\documentclass{article}
\usepackage{hyperref}
\usepackage{graphicx}
\usepackage[utf8x]{inputenc}
\usepackage[T1,T2A]{fontenc}
\usepackage[russian,english]{babel}

\usepackage{amsmath,amssymb,amsthm,amsfonts}

\newtheorem{lemma}{Lemma}

\newcommand{\Li}{\mathrm{Li}}
\newcommand{\dx}{\mathrm{dx}~}
\newcommand{\dxb}{\mathrm{d\bar x}~}
\newcommand{\dy}{\mathrm{dy}~}
\newcommand{\dz}{\mathrm{dz}}

\begin{document}
\title{Finite-size scaling of free energy in the dimer model on a hexagonal domain}

\author{A.A.~Nazarov$^{1,2}$, S.A.~Paston$^{1}$\\
{\small
  $^{1}$Department of Physics, St. Petersburg State University,} \\
{\small  Ulyanovskaya 1, 198504 St.~Petersburg, Russia}\\
\small{$^{2}$email:antonnaz@gmail.com}
}
\date{}
\maketitle

\begin{abstract}
  We consider dimer model on a hexagonal lattice. This model can be
  seen as a ``pile of cubes in the box''. The energy of configuration
  is given by the volume of the pile and the partition function is
  computed by the classical MacMahon formula or, more formally, by the
  determinant of Kasteleyn matrix. We use the expression for the
  partition function to derive the scaling behavior of free energy in
  the limit of lattice mesh tending to zero and temperature tending to
  infinity. We consider the cases of finite hexagonal domain, of
  infinite height box and coordinate-dependent Boltzmann weights. We
  obtain asymptotic expansion of free energy and discuss the
  universality and physical meaning of the expansion coefficients.
\end{abstract}

\section*{Introduction}
\label{sec:introduction}
The dimer model appeared in 1937 in an attempt to extend a statistical theory of perfect solutions
in chemistry to the case of liquid mixtures with molecules of two very distinct sizes
\cite{Fowler-1937}. The molecules were represented by the rigid tiles on a lattice and the number of
tilings was approximately estimated. But an exact computation was not accessible at the time.

In 1961, Kasteleyn \cite{P.W-1961}, Temperley and Fisher \cite{doi:10.1080/14786436108243366}
represented the partition function of the dimer model as a Pfaffian of the signed adjacency matrix
(``Kasteleyn matrix'') thus allowing the computation of the number of tilings and of free energy scaling limit.
This result was very elegantly used by Fisher to solve Ising model \cite{fisher1966dimer} and by Fan
and Wu \cite{Fan-1970} to compute free energy for a certain case of eight-vertex model.

Further studies of dimer models revealed the connection to the theory of alternating matrices
\cite{elkies1992alternating1,elkies1992alternating2}. Later, the well-known limit shape
phenomenon~\cite{vershik1977kerov} was discovered for dimer models. First, the ``Arctic circle''
theorem was proven for domino tilings of the domain in the form of ``Aztec diamond''
~\cite{1998math......1068J}. Then similar result was obtained for a hexagonal domain on the
hexagonal lattice~\cite{cohn1998shape}. Soon the connection of these results with the theory of
random matrices was established~\cite{johansson2002non}. The papers
\cite{kenyon2006dimers,kenyon2009lectures} present a detailed exposition of the limit shape
phenomenon in the dimer models.

Dimer models are the integrable lattice models of statistical physics that are now under an active
theoretical~\cite{zj2000,ferrari} and numerical investigation~\cite{ks2018}. Computation of
correlation functions is a common problem for all vertex models \cite{colomo2012approach} as well as
for dimer models. Another problem of great interest is the study of limit shapes in various cases
\cite{borodin2010q,di2018tangent}.

Configurations of dimer model on a hexagonal lattice are in one-to-one correspondence with the
configurations of five-vertex model that appears for certain choice of parameters in the well-known
six-vertex model \cite{kapitonov2012weighted,kapitonov2008five}.

Study of dimer models on various lattices and domains led to interesting connections with the
geometry of curved manifolds and with spectra of discrete and continuous Dirac and Laplace operators
\cite{kenyon2002laplacian,kenyon2000asymptotic}. Scaling limit of dimer model is proven to be
described by a Gaussian free-field theory \cite{kenyon2001dominos}, but finite-size corrections were
not considered previously. These corrections are important to close the gap between numerical
simulations and theoretical results.

We consider a particular case of the dimer model on a hexagonal domain
of hexagonal lattice, that can be seen as a ``pile of cubes in the
corner of the box''. The energy of configuration is the total number
of cubes. For this particular case we use an exact combinatorial
formula for the partition function to derive the expressions for
scaling limit of free energy and first three terms of the finite-size
corrections. We show that the first term is identically zero. The
second term which encapsulates logarithmic dependence on the mesh size
is connected with the central charge of the effective field theory and
with the geometry of limit shape. The third term has a
geometry-dependent contribution that is written explicitly in terms of
elementary functions and a universal contribution, which is constant
in all the cases that we consider.

We then consider the box of infinite height and the case of
coordinate-dependent Boltzmann weights. In these cases we are also
able to write explicit expressions for finite-size corrections. We
demonstrate the universality of the logarithmic term and the constant
contribution to the third term.

Our results are supported by numeric checks and Monte-Carlo
simulations presented in our previous publication
\cite{1742-6596-1135-1-012024}.

\section{Model definition}
\label{sec:model-definition}
The configurations of the dimer model are perfect matchings (sets of non-touching edges, covering
all the vertices) on some graph ${\cal G}$ with some choice of weights $\omega(e)$ on the edges. The
model is solvable on the bipartite graphs, i.e. the partition function can be computed if the
weights are introduced in such a way that for each face bounded by 0 mod 4 edges there is an odd
number of negative edge weights and each face with 2 mod 4 edges has an even number of of negative
edge weights. Then the signs of the edge weights form a so-called ``Kasteleyn orientation'' on graph,
the weighting is called ``Kasteleyn weighting'' \cite{kenyon2001dominos,kenyon2009lectures}.

For a bipartite graph ${\cal G}$, color the vertices black and white in such a way that all the
vertices adjacent to the black one are white. Denote by $B, W$ the sets of black and white
vertices and by $b,w$ the elements of these sets. 

The weights can be encoded as the ``Kasteleyn matrix'' -- weighted, signed adjacency matrix ${\cal K}$ with
the matrix elements ${\cal K}(w,b)$ equal to the weight of the edge $w\to b$: ${\cal K}(w,b)=\omega(w\to b)$.

Then the partition function is equal to the absolute value of the determinant of the Kasteleyn
matrix\cite{P.W-1961,doi:10.1080/14786436108243366}: 
\begin{equation}
  \label{eq:15}
  Z=\sum_{\mathrm{conf}}\prod_{e\in \mathrm{conf}}w(e)=|\det {\cal K}|.
\end{equation}

Kasteleyn matrix defines a discrete Dirac operator $D$, the action of $D$ on a function $f$ defined
on vertices is given by:
\begin{equation}
  \label{eq:16}
  (Df)(v)=\sum_{u}{\cal K}(v,u) f(u).
\end{equation}

Kenyon \cite{kenyon2002laplacian,kenyon2000asymptotic} and others \cite{sridhar2015asymptotic}
considered asymptotics of the determinants of the discrete Dirac and Laplace operators, the problem
that, as can be seen from the above, is very close to the scaling of the free energy. But the finite
size corrections to the free energy scaling were not computed. 
  
We consider coverings of the hexagonal domain on the hexagonal lattice consisting of the subsets of
lattice edges such that every vertex is the endpoint of exactly one edge.

\begin{figure}[htbp]
\center{\scalebox{0.4}{\includegraphics{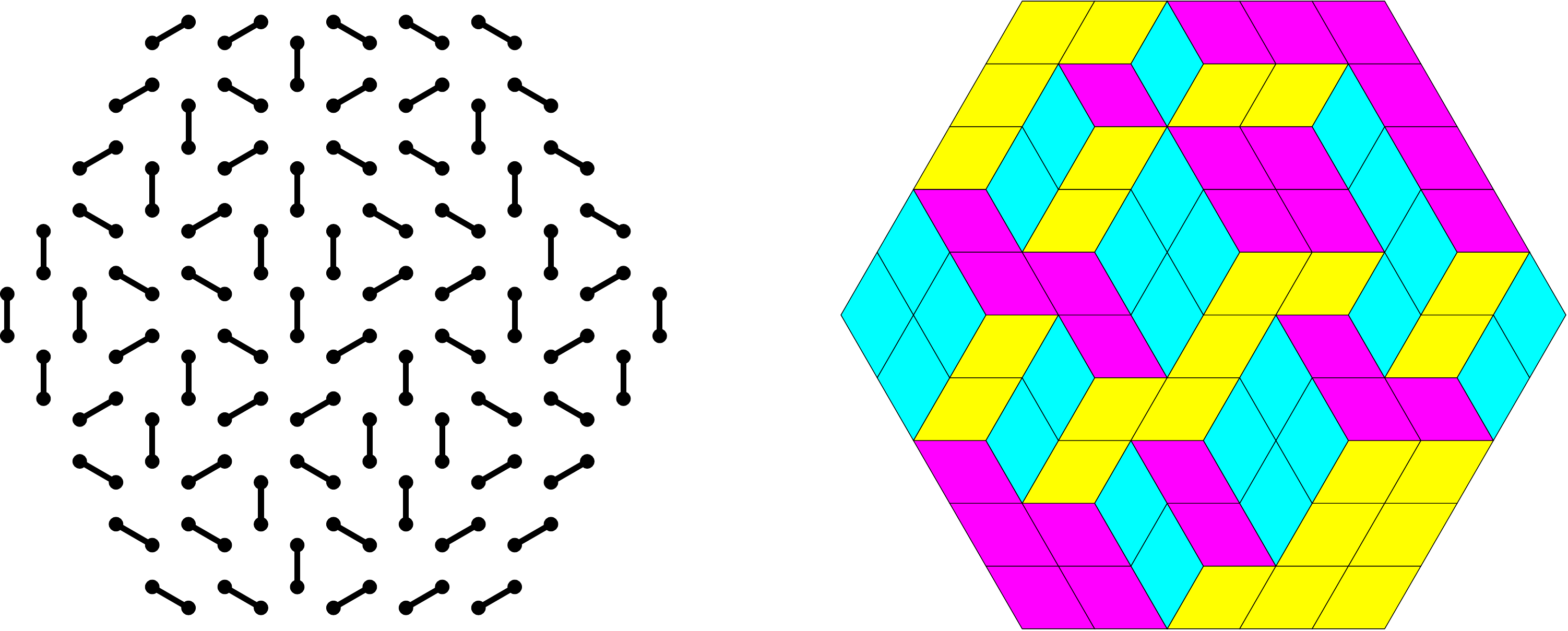}}}
\caption{\label{dhf}A configuration of dimers on the hexagonal lattice and a corresponding picture
  of ``cubes in the corner''.}
\end{figure}

We can draw a rhombus on a dual lattice around each edge in the configuration. The picture of
``cubes in the corner'' presented in the Fig.~\ref{dhf} is obtained. Let us write on the top of each
uppermost cube the height of its column of cubes. Looking at this picture from the top, we obtain a
height function defined on the rectangular domain of the square lattice. 

Let us define the sizes $M$, $N$, and $K$ of the sides of the hexagon.
The above description can be formalized by
setting the non-negative numbers up to $K$ in the boxes of the rectangular $M\times N$ table so that a value in
each box is not greater than values in the adjacent upper and left boxes
\begin{equation}
  \label{eq:1}
  h_{ij}\leq h_{i-1,j},\quad h_{ij}\leq h_{i,j-1}.
\end{equation}

The weight of a particular configuration is given by the exponent of the volume of all cubes or by a
sum of the height function values:
\begin{equation*}
  \label{eq:10}
  E[conf]=\sum_{i,j} h_{ij}=\mathrm{Volume}
\end{equation*}
We set Boltzmann constant equal to 1 and choose system of units in such a way that there is no
coupling constant in the expression for energy. Then the partition function is
\begin{equation*}
  \label{eq:14}
  Z=\sum_{conf} e^{-\frac{E[conf]}{T}}=\sum_{conf}q^{\mathrm{Volume}[conf]}, 
\end{equation*}
where $q=\exp\left(-1/T\right)$ .

For this particular case, the partition function is given by the classical Macmahon combinatorial
formula~\cite{vuletic2009generalization}
\begin{equation}
  \label{eq:12}
   Z[M,N,K,q]=\prod_{i=1}^{M}\prod_{j=1}^{N}\prod_{k=1}^{K}\frac{1-q^{i+j+k-1}}{1-q^{i+j+k-2}}.
\end{equation}

MacMahon formula is obtained for the following definition of the Kasteleyn matrix. Embed the
hexagonal lattice in the complex plane $\mathbb{C}$ in such a way that some edges are parallel to
the real line with and corresponding vertices have coordinates with integer real and imaginary
parts. Then take
\begin{eqnarray}
  \label{eq:18}
  {\cal K}(w,b)=q^{\Re w+\Im w} \quad \mathrm{if}\quad \Im w=\Im b\\
  {\cal K}(w,b)=1 \quad \mathrm{if}\quad \Im w\neq\Im b.
\end{eqnarray}

The   free energy per site is defined as$^{*}$
\footnote{For convenience we omit the factor $\frac{1}{T}$ in the usual definition of the free energy.}
\begin{equation*}
  \label{eq:17}
  f=-\frac{1}{V}\ln Z(M,N,K,q).
\end{equation*}
Here $V$ is the number of vertices, it is twice the number of dimers and twice the number of  cube faces:
\begin{equation}
  \label{eq:19}
  V=2(MN+NK+MK)=2(ab+bc+ca) \varepsilon^{-2}.
\end{equation}

We are interested in the scaling limit, combined with the thermodynamic limit, when $ T\to \infty,$
and $M,N,K\to \infty$, such that ratios $\frac{M}{T}=a,\quad \frac{N}{T}=b, \quad \frac{K}{T}=c$
remain fixed. In what follows we use $\varepsilon=\frac{1}{T}$, which can be seen as the scale of
the model, e.g. mesh size due to our choice of the system of units.

In the next section we compute the asymptotic expansion of the free energy $f$ in $\varepsilon$ and
derive $\varepsilon$-independent closed expressions for the first several coefficients in this
expansion.
  
\section{The computation of the free energy asymptotic expansion}
\label{sec:free-energy-scaling}
First we substitute MacMahon formula \eqref{eq:12} into the free energy definition \eqref{eq:17} and
obtain
\begin{multline}
  \label{eq:20}
    f=-\frac{1}{V}\ln Z =\\=- \sum_{i=1}^{M} \sum_{j=1}^{N} \sum_{k=1}^{K} \frac{1}{V}\left[
  \ln\left(1-e^{-\varepsilon (i+j+k-3)} e^{-2\varepsilon}\right) -\ln\left(1-e^{-\varepsilon
      (i+j+k-3)} e^{-\varepsilon}\right)\right].
\end{multline}
We need to notice that for a finite $\varepsilon$ and each value of $i,j,k$ the logarithms have the
form $\ln(1-x)$, where $0<x<1$. Thus we can represent the logarithms as a power series of $x$ and
obtain
\begin{equation}
  \label{eq:7}
  f=-\sum_{i=1}^{M} \sum_{j=1}^{N} \sum_{k=1}^{K}\sum_{n=1}^{\infty}
  \frac{1}{V}\frac{1}{n}\left[-e^{-n\varepsilon (i+j+k-3)} e^{-2n\varepsilon}+e^{-n\varepsilon
      (i+j+k-3)} e^{-n\varepsilon}\right]. 
\end{equation}
Here we have an absolutely convergent series, so we can change the order of summation and rewrite
the free energy as
\begin{equation}
  \label{eq:37}
  f=-\sum_{n=1}^{\infty}
  \frac{1}{V}\frac{1}{n}e^{-n\varepsilon}\left(1-e^{-n\varepsilon}\right)\sum_{i=1}^{M}
  \sum_{j=1}^{N} \sum_{k=1}^{K}e^{-n\varepsilon (i+j+k-3)} .
\end{equation}
The triple sum factorizes into the product of three sums of the type $\sum_{i=0}^{M-1}e^{-n\varepsilon
  i}$, that are just the
 sums of geometric progression. So we have $\sum_{i=0}^{M-1}e^{-n\varepsilon
  i}=\frac{1-e^{-Mn\varepsilon}}{1-e^{-n\varepsilon}}$, and obtain
\begin{equation}
  \label{eq:38}
  f=-\sum_{n=1}^{\infty}
  \frac{1}{V}\frac{1}{n}\frac{e^{-n\varepsilon}}{\left(1-e^{-n\varepsilon}\right)^{2}}
  \left(1-e^{-na}\right)\left(1-e^{-nb}\right)\left(1-e^{-nb}\right).
\end{equation}
Denote by $\chi(z)$ the function
\begin{equation}
  \label{eq:39}
  \chi(z)=e^{-z}\left(\frac{z}{1-e^{-z}}\right)^{2}, 
\end{equation}
and by $H_{n}$ the product of $\varepsilon$-independent exponents
\begin{equation}
  \label{eq:40}
  H_{n}=\left(1-e^{-na}\right)\left(1-e^{-nb}\right)\left(1-e^{-nc}\right).
\end{equation}

The function $\chi(z)$ is smooth in 0, even and has the following Taylor series:
\begin{equation}
  \label{eq:45}
  \chi(z)=1 - \frac{z^2}{12} + \frac{z^4}{240}+\mathcal{O}(z^{6}).
\end{equation}

Using the definition of the volume \eqref{eq:19} the free energy is written as
\begin{equation}
  \label{eq:41}
  f=-\frac{1}{2(ab+bc+ca)} \sum _{n=1}^{\infty} \frac{\chi(n\varepsilon)}{n^{3}} H_{n}.
\end{equation}
We are interested in the scaling behavior in $\varepsilon$ when $\varepsilon\to 0$. The function $f$
is regular in $\varepsilon$ for $\varepsilon>0$. So we need to
compute  $\left.f\right|_{\varepsilon\to 0}$ and the derivatives $\left.\frac{\partial f}{\partial
  \varepsilon}\right|_{\varepsilon\to 0}$, $\left.\frac{\partial^{2} f}{\partial
  \varepsilon^{2}}\right|_{\varepsilon\to 0}$.

For $\left.f\right|_{\varepsilon\to 0}$ we have $\chi(0)=1$ and the sum $\sum_{n=1}^{\infty}
\frac{H_{n}}{n^{3}}$ is just a sum of  polylogarithms
$\mathrm{Li}_{s}(z)=\sum_{n=1}^{\infty}\frac{z^{n}}{n^{s}}$ of third order:
\begin{multline}
  \label{eq:42}
  \left.f\right|_{\varepsilon\to 0} =\frac{1}{2(ab+bc+ca)}\left[\Li_{3}(e^{-a})+\Li_{3}(e^{-b})+\Li_{3}(e^{-c})-
    \Li_{3}(e^{-a-b})\right.\\
  \left.-\Li_{3}(e^{-b-c})-    \Li_{3}(e^{-a-c})+    \Li_{3}(e^{-a-b-c})-\zeta(3)\right].
\end{multline}
Here Riemann zeta function appears as a particular value of polylogarithm $\Li_{3}(1)=\zeta(3)$.

The derivative of the function $\chi(z)$ is zero, $\chi'(0)=0$, and the series
$\sum_{n=1}^{\infty} \frac{\chi'(n\varepsilon)}{n^{2}}H_{n}$ are convergent for a finite
$\varepsilon$, thus we have for a derivative
\begin{equation}
  \label{eq:43}
\left.\frac{\partial f}{\partial \varepsilon}\right|_{\varepsilon\to 0}=0.
\end{equation}

Now let us compute the second derivative
\begin{equation}
  \label{eq:44}
\frac{\partial^{2} f}{\partial
  \varepsilon^{2}}=-\frac{1}{2(ab+bc+ca)} \sum _{n=1}^{\infty} \frac{\chi''(n\varepsilon)}{n} H_{n}.  
\end{equation}
Second derivative of $\chi$ is finite $\chi''(0)=-\frac{1}{6}$. But there is a difficulty with the
limit $\varepsilon\to 0$ in this expression due to the poor convergence of the series. Let us
rewrite the second derivative as a sum of two series:
\begin{equation}
  \label{eq:46}
\frac{\partial^{2} f}{\partial
  \varepsilon^{2}}=-\frac{1}{2(ab+bc+ca)} \left(\sum _{n=1}^{\infty} \frac{\chi''(n\varepsilon)}{n}+
  \sum_{n=1}^{\infty} \frac{\chi''(n\varepsilon)}{n}(H_{n}-1)\right).    
\end{equation}
The second sum is now convergent, since $H_{n}-1$ decays exponentially in $n$. It can be presented
as a combination of series for logarithms:
\begin{multline}
  \label{eq:47}
  \chi''(0)\sum_{n=1}^{\infty} \frac{H_{n}-1}{n}=\\=\chi''(0)\sum_{n=1}^{\infty}\frac{1}{n}\left(-e^{-a
      n-b n-c n}+e^{-a n-b n}+e^{-a n-c n}+e^{-b n-c n} - e^{-a n}-e^{-b n}-e^{-c n}\right)=\\
  =\frac{1}{6}\ln\left(\frac{(e^{a}-1)(e^{b}-1)(e^{c}-1)(e^{a+b+c}-1)}{(e^{a+b}-1)(e^{b+c}-1)(e^{a+c}-1)}\right).
\end{multline}

First sum in the expression \eqref{eq:46} can be expressed as follows. First, consider second
derivative of $\chi(z)$, separate an exponent and denote what is left by $\xi(z)$:
\begin{equation}
  \label{eq:48}
  \chi''(z)=e^{-z}\frac{e^{2 z} \left(z^2+4 e^z
   \left(z^2-1\right)+e^{2 z} \left(z^2-4
   z+2\right)+4
   z+2\right)}{\left(e^z-1\right)^4}=e^{-z}\xi(z).
\end{equation}
The sum is then rewritten
\begin{equation}
  \label{eq:49}
  \sum _{n=1}^{\infty} \frac{\chi''(n\varepsilon)}{n}=
  \sum_{n=1}^{\infty}\frac{e^{-n\varepsilon}}{n}\xi(n\varepsilon) =
  \sum_{n=1}^{\infty}\frac{e^{-n\varepsilon}}{n}\xi(0)+  \varepsilon\sum_{n=1}^{\infty}e^{-n\varepsilon}\frac{\xi(n\varepsilon)-\xi(0)}{n\varepsilon}.
\end{equation}
The first term is series for the logarithm $\ln\left(1-e^{-\varepsilon}\right)\approx \ln
\varepsilon$. 
Denote by $Q(z)$ the difference of the function $\xi$ value in $z$ and zero, divided by z:
\begin{equation}
  \label{eq:85}
  Q(z)=\frac{\xi(z)-\xi(0)}{z}.
\end{equation}
The function $Q(z)$ is analytic in $z$.

Then the second term can approximated by the integral:
\begin{equation}
  \label{eq:50}
  \varepsilon\sum_{n=1}^{\infty}e^{-n\varepsilon}Q(n\varepsilon)=\varepsilon\sum_{n=1}^{\infty}e^{-n\varepsilon}\frac{\xi(n\varepsilon)-\xi(0)}{n\varepsilon}\approx
  \int_{0}^{\infty} e^{-z}\frac{\xi(z)-\xi(0)}{z} dz=\int_{0}^{\infty}e^{-z}Q(z)\dz.
\end{equation}
\begin{equation}
  \label{eq:51}
  \frac{\partial^{2} f}{\partial
  \varepsilon^{2}}\approx-\frac{1}{2(ab+bc+ca)}
\left[\frac{1}{6}\ln\left(\frac{(e^{a}-1)(e^{b}-1)(e^{c}-1)(e^{a+b+c}-1)}{(e^{a+b}-1)(e^{b+c}-1)(e^{a+c}-1)}\right)-\frac{1}{6}\ln \varepsilon+\int_{0}^{\infty} e^{-z}\frac{\xi(z)-\xi(0)}{z} dz\right].
\end{equation}
Since we have logarithmic term in the second derivative, we have the behavior of free energy of the form
\begin{equation}
  \label{eq:26}
  f\approx f_{0}+f_{1}\varepsilon +f_{2}\varepsilon^{2}\ln\varepsilon+ f_{3}\varepsilon^{2}+\mathcal{O}(\varepsilon^{3}),
\end{equation}
as expected in general in two-dimensional critical systems \cite{cardy1988finite}. Taking the
derivatives of such expression by $\varepsilon$, we get $f_{0}=f(0)$,
$f_{1}=\left.\frac{\partial f}{\partial \varepsilon}\right|_{\varepsilon\to 0}$,
$\frac{\partial^{2} f}{\partial \varepsilon^{2}}=f_{2}(2\ln\varepsilon+3)+f_{3}$. 

Thus we have the coefficients
\begin{multline}
  \label{eq:27}
  f_{0}=\frac{1}{2(ab+bc+ca)}\left[\Li_{3}(e^{-a})+\Li_{3}(e^{-b})+\Li_{3}(e^{-c})-
    \Li_{3}(e^{-a-b})\right.\\
  \left.-\Li_{3}(e^{-b-c})-    \Li_{3}(e^{-a-c})+    \Li_{3}(e^{-a-b-c})-\zeta(3)\right],
\end{multline}
\begin{equation}
  \label{eq:28}
  f_{1}=0,
\end{equation}
\begin{equation}
  \label{eq:30}
  f_{2}=-\frac{1}{2(ab+bc+ca)}\frac{1}{12},
\end{equation}
and
\begin{equation}
  \label{eq:29}
  f_{3}=-\frac{1}{2(ab+bc+ca)}\left[\frac{1}{12}\ln\left(\frac{(e^{a}-1)(e^{b}-1)(e^{c}-1)(e^{a+b+c}-1)}{(e^{a+b}-1)(e^{b+c}-1)(e^{a+c}-1)}\right)-\frac{1}{8}+ \frac{1}{2}\int_{0}^{\infty} e^{-z}Q(z) dz
    \right].
\end{equation}

\section{Limit shape and physical meaning of the expansion coefficients}
\label{sec:accur-expans-phys}

In the paper \cite{1742-6596-1135-1-012024} we have presented results of Monte Carlo simulations using
Metropolis and Wang-Landau algorithms that support the form of expansion (\ref{eq:26}). In
particular we got $f_{1}=0.04\pm 0.04$ which is consistent with $f_{1}=0$. 

The dimers in the scaling limit are in general described by a
free-fermion theory \cite{dijkgraaf2009dimer}, but on a hexagonal
lattice (and on the graphs that admit Kasteleyn orientation) the model
is reformulated in terms of height function, that can be seen as a
bosonization of the theory \cite{gogolin2004bosonization}.

In the scaling limit $\varepsilon\to 0$ so called ``limit shape
phenomenon''~\cite{1998math......1068J,cohn1998shape} appears in the
dimer model. The areas around the corners of the domain are ``frozen''
with height function values being fixed. An analytical ``Arctic
curve'' delimits frozen regions from the region where the behavior is
described by the effective free field theory
\cite{kenyon2009lectures,kenyon2008height,kenyon2006dimers}. Moreover,
the height function in the scaling limit tends to certain analytical
surface and fluctuations around this ``limit shape'' are described by
a gaussian free field theory in a curved background of ``limit
shape''. A confomal map can be used to send the internal region of the
``Arctic curve'' onto the upper half-plane.

The scaling behavior (\ref{eq:26}) of the logarithm of the partition function is generic in two-dimensional
models \cite{cardy1988finite}.

Thus we can interpret first two terms $f_{0}$ and $f_{1}$ as a bulk
and boundary free energies in the corresponding field theory. The term
$f_1$ corresponds to the contribution of the ``limit shape''. Since we
have a deterministic frozen boundary, we naturally have $f_{1}=0$.

The term proportional to the logarithm of the scale $\varepsilon$ is also universal
\cite{cardy1988finite} and should appear in all two-dimensional theories with boundary. In the paper
\cite{cardy1988finite} Cardy and Peschel argued that on a smooth manifold of a characteristic length $L$ with a smooth
boundary such a term must have the following form:
\begin{equation}
  \label{eq:31}
  \delta F = -\frac{1}{6}{\bf c} \chi \ln L,
\end{equation}
where ${\bf c}$ is central charge of the effective field theory and
$\chi$ is the Euler characteristic of the manifold $\chi=2-2 h-b$,
where $h$ is the number of handles and $b$ is the number of
boundaries. Naive use of the formula \eqref{eq:31} with our result
$\delta F=\frac{1}{12}$ would give ${\bf c}=\frac{1}{2}$, but it is
not correct, since we need to consider the field theory in the curved
background.

In the case of a corner singularity on the boundary with the angle
$\Theta$ the logarithmic contribution to the free energy is changed to
\begin{equation}
  \label{eq:86}
  \delta F = \frac{{\bf c}}{24}\left(\frac{\Theta}{\pi}-\frac{\pi}{\Theta}\right) \ln L.
\end{equation}
This formula was recently used by N. Allegra \cite{allegra2015exact} to show that the central charge
in the dimer model on the square lattice with a corner monomer is ${\bf c}=1$, contrary to some
previous claims of ${\bf c}=-2$.

In the case of manifold with conic singularity with the semi-angle
$\Theta$, the logarithmic contribution presented in \cite{cardy1988finite} is:
\begin{equation}
  \label{eq:91}
  \delta F = \frac{{\bf c}}{24}\left(\frac{\Theta}{\pi}-\frac{4\pi}{\Theta}\right) \ln L.
\end{equation}

In our case we interpret $2(ab+bc+ca)$ as a volume of the domain, thus
\begin{equation}
  \label{eq:33}
  \delta F = \left(2(ab+bc+ca)\right) f_{2}\ln\varepsilon.
\end{equation}

Using Cardy-Peschel formulas, presented above, $\delta F$ can be
computed from value of the central charge for Gaussian free field
${\bf c}=1$, the curvature of metric, induced by the limit shape and
conformal map from the internal region of the ``Arctic curve'' to the
upper half-plane. Such a computation can be used to check our result
on $f_2$ but is quite lengthy and will be presented in a separate
publication.

The last term $f_{3}$ depends only on the shape of the domain through
$a,b,c$ with a universal contribution that is equal to the integral of
a function $e^{-z}\frac{\xi(z)-\xi(0)}{z}$. In the next sections we
show that this constant appears due to the treatment of logarithmic
divergency even for the coordinate-dependent Boltzmann weights. 

\section{Numeric checks}
\label{sec:numeric-checks}
To check our result \eqref{eq:26} numerically, we first need to evaluate the integral in $f_{3}$.
The function $e^{-z}\frac{\xi(z)-\xi(0)}{z}$ is smooth and but some care is required when taking the
limit $z\to 0$ numerically. Evaluating the integral numerically with Mathematica we get
\begin{equation}
  \label{eq:2}
  \int_{0}^{\infty}e^{-z}Q(z)\dz=\int_{0}^{\infty}e^{-z}\frac{\xi(z)-\xi(0)}{z} dz = -0.080842.
\end{equation}

\begin{figure}[htbp]
  \includegraphics[scale=0.9]{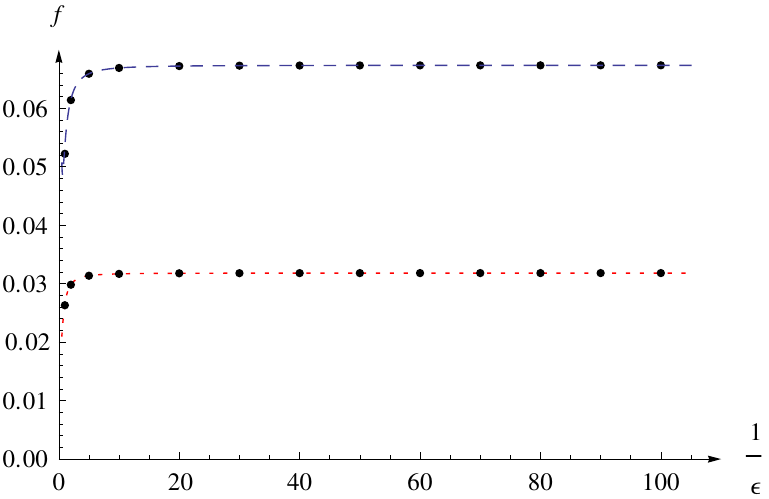}
  \includegraphics{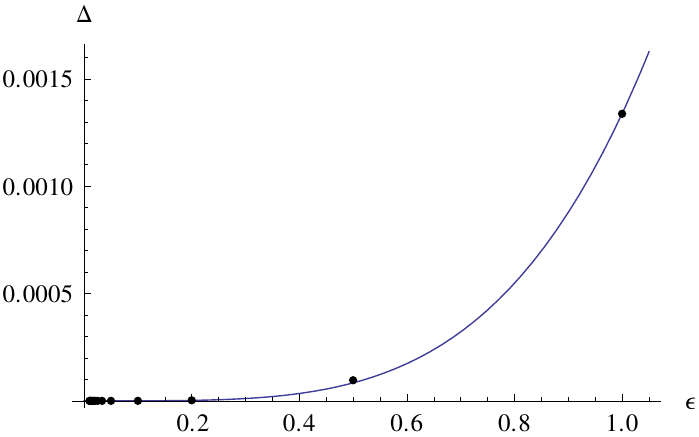}
  \caption{\label{fig:approx-acc} {\it On the left:} Dependence of free energy on $1/\varepsilon$,
    top -- $a=b=c=1$, bottom -- $a=3, b=2, c=1$. Exact values are shown by solid dots.
    Approximations with formula \eqref{eq:26} are shown as blue dashed and red dotted lines. {\it On
      the right:} Dependence of approximation inaccuracy on $\varepsilon$, solid line is fit by $b\varepsilon^{4}$.}
\end{figure}

The asymptotic expansion \eqref{eq:26} gives very good approximation for free energy. In Figure
\ref{fig:approx-acc} we present exact and approximate values for $f$ and, on the right panel,
demonstrate that inaccuracy is actually $\mathcal{O}(\varepsilon^{4})$. 

\section{Box of infinite height}
\label{sec:gener-other-doma}

Consider the case of cubes in the box of infinite height. In this case $K=\infty$ and $M,N$ are
finite. Now the hexagonal domain of dimer configurations has side of infinite length and total
number of dimers is infinite, so it is more natural to set the volume of the system $V=MN$.  

Putting $K=\infty$ in the partition function, we see that all terms with $k$ are cancelled and we have
\begin{equation}
  \label{eq:3}
  Z[M,N,q]=\prod_{i=1}^{M}\prod_{j=1}^{N}\frac{1}{1-q^{i+j-1}}.
\end{equation}

The computation of free energy is almost exactly the same as in finite box case. We expand the
logarithm in the free energy and then factorize the sum:
\begin{multline}
  \label{eq:4}
  f=-\frac{1}{MN}\sum_{i=1}^{M}\sum_{j=1}^{N}\ln\left(1-e^{-\varepsilon(i+j-2)}e^{-\varepsilon}\right)=\\
  =\sum_{n=1}^{\infty}\frac{\varepsilon^{2}}{ab}\frac{1}{n}e^{-n\varepsilon}\left(\sum_{i=1}^{M}e^{-n\varepsilon(i-1)}\right)
  \left(\sum_{j=1}^{N}e^{-n\varepsilon(j-1)}\right).
\end{multline}
Computing the sums of geometric progressions we obtain almost the same expression as \eqref{eq:38},
except that $H_{n}$ now has two multipliers $H_{n} =\left(1-e^{-na}\right)\left(1-e^{-nb}\right)$:
\begin{equation}
  \label{eq:5}
  f=-\frac{1}{ab} \sum_{n=1}^{\infty}\frac{\chi(n\varepsilon)}{n^{3}} H_{n}.
\end{equation}
The function $\chi(z)=e^{-z}\left(\frac{z}{1-e^{-z}}\right)^{2}$ is the same as in \eqref{eq:39}.

Doing the same computations as in section \ref{sec:free-energy-scaling}, we obtain following results
for the coefficients $f_{0}, f_{1}, f_{2}, f_{3}$:
\begin{equation}
  \label{eq:6}
  f_{0}=\frac{1}{ab}\left[\Li_{3}(e^{-a})+\Li_{3}(e^{-b})-
    \Li_{3}(e^{-a-b})-\zeta(3)\right],
\end{equation}
\begin{equation}
  \label{eq:8}
  f_{1}=0,
\end{equation}
\begin{equation}
  \label{eq:9}
  f_{2}=\frac{1}{ab}\frac{1}{12},
\end{equation}
\begin{equation}
  \label{eq:11}
  f_{3}=-\frac{1}{ab}\left[\frac{1}{12}\ln\left(\frac{(e^{a}-1)(e^{b}-1)}{(e^{a+b}-1)}\right)-\frac{1}{8}+ \frac{1}{2}\int_{0}^{\infty} e^{-z}\frac{\xi(z)-\xi(0)}{z} dz
    \right].
\end{equation}

\begin{figure}[htbp]
  \includegraphics[scale=0.8]{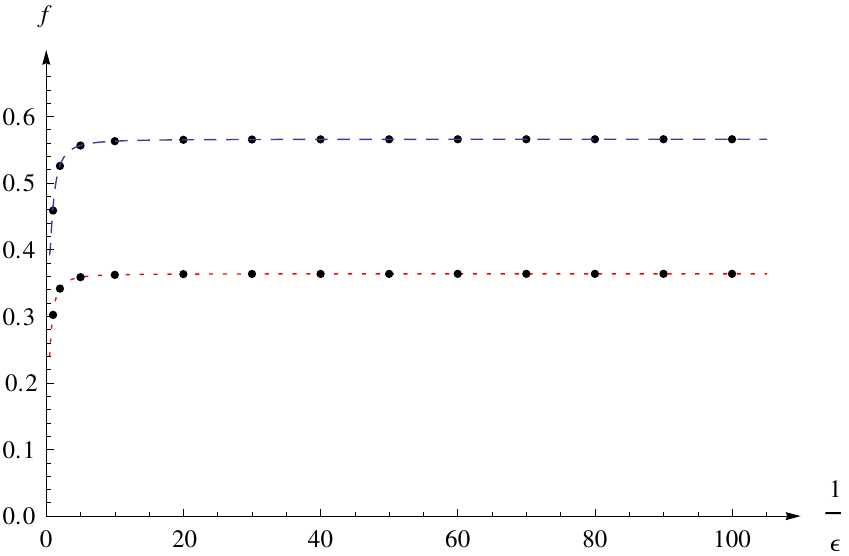}
  \includegraphics[scale=0.9]{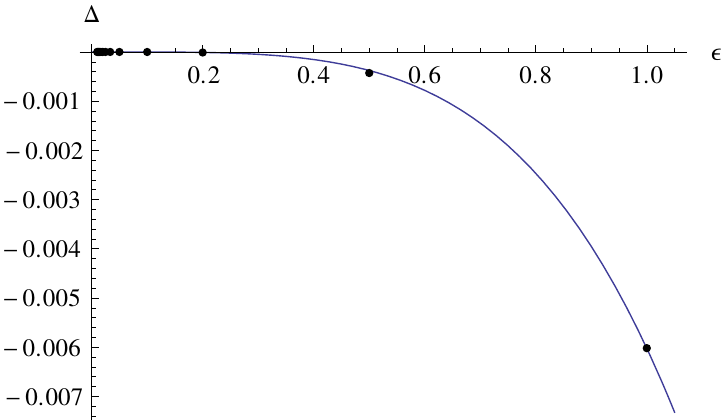}
  \caption{\label{fig:approx-acc-2d} {\it On the left:} Dependence of free energy on $1/\varepsilon$
    in the infinite height box,
    top -- $a=b=1$, bottom -- $a=2, b=1$. Exact values are shown by solid dots.
    Approximations with formula \eqref{eq:26} are shown as blue dashed and red dotted lines. {\it On
      the right:} Dependence of approximation inaccuracy on $\varepsilon$, solid line is fit by $b\varepsilon^{4}$.}
\end{figure}

Again in the Fig. \ref{fig:approx-acc-2d} we see good numerical agreement between the asymptotic
expansion \eqref{eq:26} and the exact formula \eqref{eq:4}.

\section{Box of infinite height with coordinate-dependent weight}
\label{sec:box-infinite-height}

\begin{figure}[htbp]
  \includegraphics[scale=0.5]{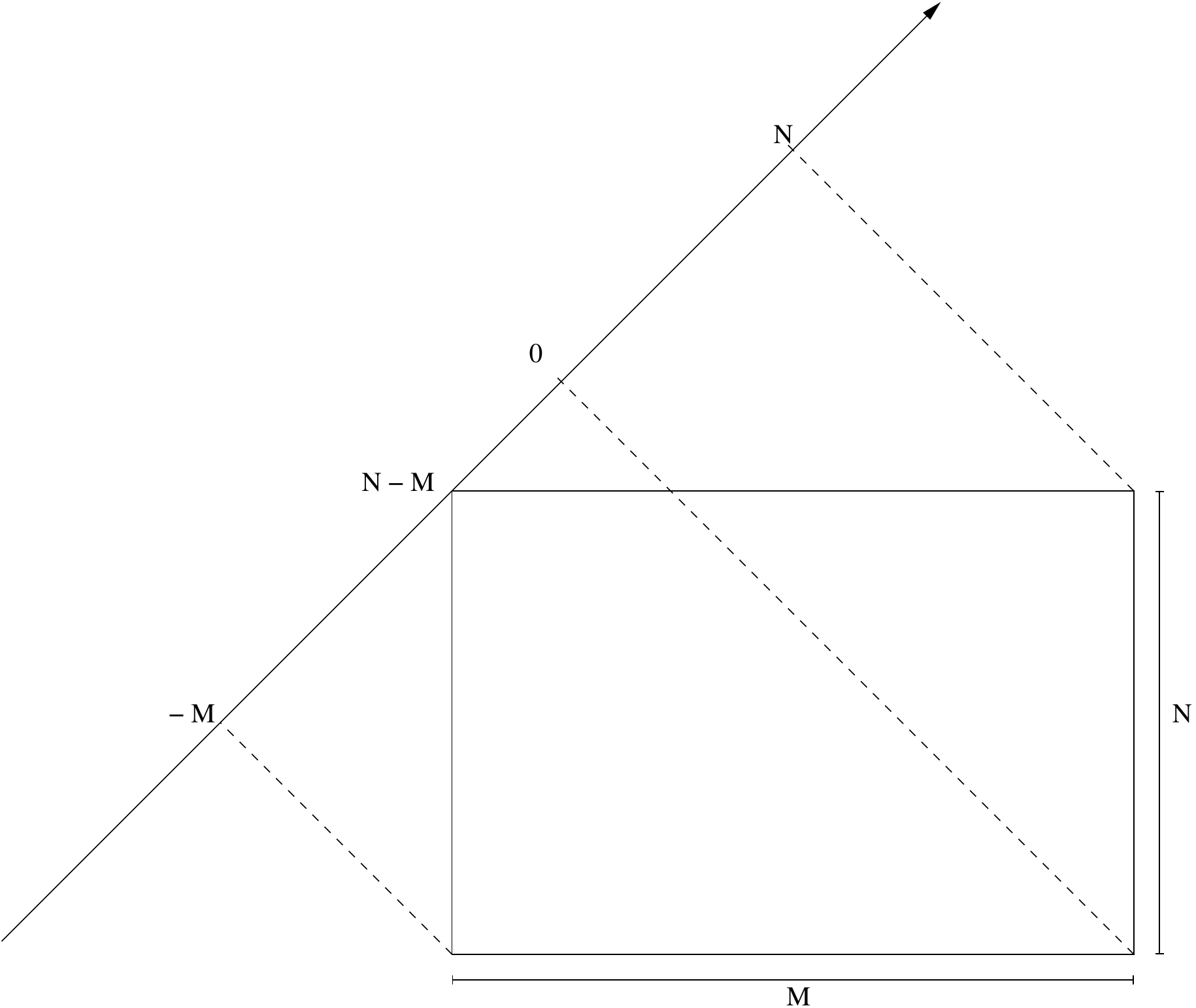}
  \caption{\label{fig:box-t-coords} Coordinates of slices.}
\end{figure}

Now let us consider the generalization to the case when $q$ depends upon the coordinate. It's usually
done by varying $q$ on diagonal slices numbered by $t$. We need proper setup to consider the scaling
limit. Let $\varphi(t)$ be a continuous bounded smooth function for $t\in\left[-a,b\right]$. Denote by
$q_{i}=e^{-\varepsilon \varphi\left(i\varepsilon\right)}$ for $i=(-M+1)\dots N-1$. We index diagonal slices in
such a way, that top left corner of the $M\times N$ box has index $N-M$, top right corner -- $N$,
bottom left -- $-M$ and bottom right -- $0$. (See Fig. \ref{fig:box-t-coords} ). The slices with the
indexes $N$ and $-M$ are empty so we set $q_{N}=q_{-M}=0$. There can be at most $N+M-1$ non-empty
slices. 

The partition function for this and more general case of arbitrary inner shape of the box was
presented in the paper \cite{okounkov2007random}. In this case it contains products of weights
$q_{i}$ for slices with negative and positive indexes:
\begin{equation}
  \label{eq:13}
  Z=\prod_{i=0}^{N-1}\prod_{j=0}^{M-1} \frac{1}{1-q_{N-M}^{-1}\left(\prod_{k=0}^{i}q_{N-M-k}\right)\left(\prod_{l=0}^{j}q_{N-M+l}\right)}.
\end{equation}
(see Theorem 2 in \cite{okounkov2007random}).
If we set $q_{i}=q$, we recover formula \eqref{eq:3}.

Now let us again consider free energy dependence on $\varepsilon$:
\begin{equation}
  \label{eq:21}
  f=-\frac{\varepsilon^{2}}{ab}\sum_{i=0}^{N-1}\sum_{j=0}^{M-1}\ln
  \left(1-e^{\varepsilon \varphi(b-a)}e^{-\varepsilon\sum_{k=0}^{i}\varphi\left(b-a-k\varepsilon\right)}
    e^{-\varepsilon\sum_{l=0}^{j}\varphi\left(b-a+l\varepsilon\right)} \right).
\end{equation}

The terms $f_{0}$  in the asymptotic expansion \eqref{eq:26} can be easily obtained from
this expression. To do so we need to approximate the sums by the integrals, using
$\sum_{k=0}^{i}\varepsilon\varphi(b-a-k\varepsilon)\approx
\int_{0}^{i\varepsilon}\varphi(b-a-x)\dx$
and similar approximation for the outer sums:
\begin{multline}
  \label{eq:74}
  f_{0}=-\frac{1}{ab}\int_{0}^{b}\int_{0}^{a}\ln
  \left(1-e^{-\int_{0}^{y}\varphi\left(b-a-x\right)\dx}
    e^{-\int_{0}^{z}\varphi\left(b-a+\bar x\right)\dxb} \right)\dz\dy=\\
  =-\frac{1}{ab}\int_{0}^{b}\int_{0}^{a}\ln
  \left(1-e^{-\int_{-y}^{z}\varphi\left(b-a+x\right)\dx} \right)\dz\dy.
\end{multline}
Substituting in this expression $\varphi(x)\equiv 1$ we immediately get formula \eqref{eq:6}, since
polylogarithms are integrals of logarithms. Although the logarithm has a divergency in the point
$y=z=0$, it disappears after double integration.

To compute the term $f_{1}$ we can use the following simple Lemma:
\begin{lemma}
  Let $a=M\varepsilon$ and $b=N\varepsilon$.
  For an integrable almost everywhere analytic function $f$ and analytic function $\varphi$ the sum is approximated by
  an integral with the corrections  of order $o(\varepsilon)$:
  \begin{equation}
    \label{eq:75}
    \sum_{i=0}^{M-1}\sum_{j=0}^{N-1}\varepsilon^{2}f\left(\varepsilon\sum_{k=-i}^{j}\varphi(k\varepsilon)\right)=
    \int_{0}^{a}\int_{0}^{b}f\left(\int_{-y}^{z}\varphi(x)\dx\right)\dz\;\dy+o(\varepsilon).
  \end{equation}
\begin{proof}
  Use Euler-Maclaurin formula to approximate the sums by the integrals with corrections. It is
  enough to consider linear in $\varepsilon$ terms. First use
  it for the sum in the argument of function $f$:
  \begin{equation}
    \label{eq:76}
    \sum_{i=0}^{M-1}\sum_{j=0}^{N-1}\varepsilon^{2}f\left(\varepsilon\sum_{k=-i}^{j}\varphi(k\varepsilon)\right)=
    \sum_{i=0}^{M-1}\sum_{j=0}^{N-1}\varepsilon^{2}f\left(\int_{-i\varepsilon}^{j\varepsilon}\varphi(x)\dx+\frac{\varepsilon}{2}\left(\varphi(j\varepsilon)-\varphi(-i\varepsilon)\right)+\mathcal{O}(\varepsilon^{2})\right).
  \end{equation}
  For a double sum the following simple analogue of Euler-Maclaurin formula can be used. 
Let $G$ be an analytic function in the rectangle $[0,a]\times[0,b]$, then
\begin{equation}
  \label{eq:77}
  \int_{0}^{a} \int_{0}^{b}G(y,z) \dz\; \dy\;
  \approx\sum_{i=0}^{M-1}\sum_{j=0}^{N-1}\left\{\varepsilon^{2}G\left(i\varepsilon,j\varepsilon\right)+
    \frac{\varepsilon^{3}}{2}\left((\partial_{y}+\partial_{z})G\right)(i\varepsilon,j\varepsilon)\right\}+o(\varepsilon^{2}).
\end{equation}
This formula can be easily derived by dividing the volume of integration into squares with side
$\varepsilon$ and substituting Taylor series for $G$ into integrals.

Denote by $G_{\varepsilon}(y,z)$ the function of two arguments that appear in the double sum in \eqref{eq:76}:
\begin{equation}
  \label{eq:78}
  G_{\varepsilon}(y,z)=f\left(\int_{-y}^{z}\varphi(x)\dx+\frac{\varepsilon}{2}(\varphi(z)-\varphi(-y))+o(\varepsilon)\right).
\end{equation}
Use formula \eqref{eq:77} to express the double sum in \eqref{eq:76} as the integral plus correction:
\begin{equation}
  \label{eq:79}
  \sum_{i=0}^{M-1}\sum_{j=0}^{N-1}\varepsilon^{2}G_{\varepsilon}(i\varepsilon,j\varepsilon)=\int_{0}^{a}\int_{0}^{b}
  G_{\varepsilon}(y,z)\dz\;\dy-
  \frac{\varepsilon^{3}}{2}
  \sum_{i=0}^{M-1}\sum_{j=0}^{N-1}\left((\partial_{y}+\partial_{z})G_{\varepsilon}\right)(i\varepsilon,j\varepsilon)+o(\varepsilon^{3}).
\end{equation}
Now expand the function $G_{\varepsilon}(y,z)$ under integral into Taylor series in $\varepsilon$:
\begin{equation}
  \label{eq:80}
  \int_{0}^{a}\int_{0}^{b}G_{\varepsilon}(y,z)\dz\;\dy=
  \int_{0}^{a}\int_{0}^{b}G_{0}(y,z)\dz\;\dy+\int_{0}^{a}\int_{0}^{b}\frac{\varepsilon}{2}(\varphi(z)-\varphi(-y))\cdot
  f'\left(\int_{-y}^{z}\varphi(x)\dx\right)\dz\;\dy+o(\varepsilon). 
\end{equation}
On the other hand, the derivatives of $G_{\varepsilon}$ in $y,z$ are computed as
\begin{eqnarray}
  \label{eq:81}
  \partial_{z}G_{\varepsilon}(y,z)=f'\left(\int_{-y}^{z}\varphi(x)\dx+\frac{\varepsilon}{2}(\varphi(z)-\varphi(-y))+o(\varepsilon)\right)\varphi(z)+o(\varepsilon)\\ 
  \partial_{y}G_{\varepsilon}(y,z)=-f'\left(\int_{-y}^{z}\varphi(x)\dx+\frac{\varepsilon}{2}(\varphi(z)-\varphi(-y))+o(\varepsilon)\right)\varphi(-y)+o(\varepsilon).
\end{eqnarray}
Substituting equations \eqref{eq:80},\eqref{eq:81} into \eqref{eq:79} and using formula
\eqref{eq:77} to change the double sum of the derivatives of $G_{\varepsilon}$ into double integral,
we see that the linear correction in $\varepsilon$ is cancelled. 

\end{proof}
\end{lemma}

Using this lemma for our free energy density $f$, which is analytic and $\varphi$ which is smooth,
we instantly see that 
\begin{equation}
  \label{eq:82}
  f_{1}=0.
\end{equation}

Next terms in the asymptotic expansion require more care, since we need to single out the
contribution of the order $\varepsilon^{2}\ln\varepsilon$.

Thus, to derive the expressions for higher order terms of the expansion \eqref{eq:26} we need to
expand the logarithm, change the order of summation and rewrite the free energy in the same way as
we did before:
\begin{equation}
  \label{eq:22}
   f=\sum_{n=1}^{\infty}\frac{\varepsilon^{2}}{ab}\frac{1}{n}e^{n\varepsilon
     \varphi(b-a)}\sum_{i=0}^{N-1}\sum_{j=0}^{M-1} e^{-n\varepsilon\sum_{k=0}^{i}\varphi\left(b-a-k\varepsilon\right)}
    e^{-n\varepsilon\sum_{l=0}^{j}\varphi\left(b-a+l\varepsilon\right)}.
\end{equation}
Now we again can factorize double sum into product of two sums:
\begin{equation}
  \label{eq:23}
    f=\sum_{n=1}^{\infty}\frac{\varepsilon^{2}}{ab}\frac{1}{n}e^{-n\varepsilon
     \varphi(b-a)}\left(\sum_{i=0}^{N-1}
     e^{-n\varepsilon\sum_{k=1}^{i}\varphi\left(b-a-k\varepsilon\right)}\right)
   \left(\sum_{j=0}^{M-1}e^{-n\varepsilon\sum_{l=1}^{j}\varphi\left(b-a+l\varepsilon\right)}\right),
 \end{equation}
where we assume that $\sum_{k=1}^{0}\varphi(b-a-k\varepsilon)=0$ and
$\sum_{l=1}^{0}\varphi(b-a+l\varepsilon)=0$. Note that for $\varphi(t)\equiv 1$ we recover formula
\eqref{eq:4}.

Introduce following notations:
\begin{equation}
  \label{eq:52}
  r_{\mp}(i\varepsilon)\equiv\varepsilon\sum_{k=1}^{i}\varphi(b-a\mp k\varepsilon).
\end{equation}
The functions $r_{\mp}$ can be approximated using Euler-Maclaurin formula as:
\begin{equation}
  \label{eq:53}
  r_{\mp}(x)\approx\int_{0}^{x}\varphi(b-a\mp y) \dy +\frac{\varepsilon}{2}\left(\varphi(b-a\mp
    x)-\varphi(b-a)\right)\mp \frac{\varepsilon^{2}}{12}\left(\varphi'(b-a\mp x)-\varphi'(b-a)\right).
\end{equation}
Denote by $\Psi_{n}^{\mp}$ the following functions:
\begin{equation}
  \label{eq:54}
  \Psi^{\mp}_{n}(x)\equiv e^{-nr_{\mp}(x)}-e^{-nxr_{\mp}'(0)}.
\end{equation}
These functions are analytic in $\varepsilon$, moreover, the higher derivatives of $\Psi^{\mp}_{n}$
grow in $n$ not too rapidly. Thus, the sums of the series written in terms of these functions and
their derivatives converge, as we will see below.

With these notations we can write the free energy as
\begin{multline}
  \label{eq:55}
  f=\frac{1}{ab}\sum_{n=1}^{\infty}\frac{1}{n}e^{-n\varepsilon\varphi(b-a)}\left[\varepsilon^{2}\sum_{i=0}^{N-1}e^{-n\varepsilon
    i r_{-}'(0)}\sum_{j=0}^{M-1}e^{-n\varepsilon j
    r_{+}'(0)}+\varepsilon\sum_{i=0}^{N-1}e^{-n\varepsilon i
    r_{-}'(0)}\varepsilon\sum_{j=0}^{M-1}\Psi^{+}_{n}(j\varepsilon)+\right.\\
\left. +\varepsilon\sum_{i=0}^{N-1}\Psi^{-}_{n}(i\varepsilon) \cdot\varepsilon\sum_{j=0}^{M-1}e^{-n\varepsilon
  j r_{+}'(0)}+\varepsilon\sum_{i=0}^{N-1}\Psi^{-}_{n}(i\varepsilon)
\cdot\varepsilon\sum_{j=0}^{M-1}\Psi^{+}_{n}(j\varepsilon)\right]=\\
=f^{(0)}+f^{+}+f^{-}+f^{\times}.
\end{multline}
The first term $f^{(0)}$ can be computed explicitly as
\begin{equation}
  \label{eq:56}
  f^{(0)}=\frac{1}{ab}\sum_{n=1}^{\infty}\frac{1}{n} e^{-n\varepsilon
    \varphi(b-a)}\varepsilon^{2}\frac{\left(1-e^{-n b r_{-}'(0)}\right)\left(1-e^{-n a
        r_{+}'(0)}\right)}{\left(1-e^{-n\varepsilon r_{-}'(0)}\right)\left(1-e^{-n\varepsilon r_{+}'(0)}\right)}.
\end{equation}

Denote by $H_{n}(\varepsilon)$ the numerator of this expression:
\begin{equation}
  \label{eq:57}
  H_{n}(\varepsilon)=\left(1-e^{-n b r_{-}'(0)}\right)\left(1-e^{-n a r_{+}'(0)}\right).
\end{equation}
Note that
\begin{equation}
  \label{eq:58}
  r_{\mp}'(0)=\varphi(b-a)\mp \frac{\varepsilon}{2} \varphi'(b-a)+\frac{\varepsilon^{2}}{12} \varphi''(b-a)+\mathcal{O}(\varepsilon^{3}).
\end{equation}
can be rewritten as
\begin{equation}
  \label{eq:59}
  r_{\mp}'(0)=\varphi(b-a)\left(1+\alpha_{\mp}(\varepsilon)\right), 
\end{equation}
where
\begin{equation}
  \label{eq:60}
  \alpha_{\mp}(\varepsilon)=\mp
  \frac{\varepsilon}{2}\frac{\varphi'(b-a)}{\varphi(b-a)}+\frac{\varepsilon^{2}}{12} \frac{\varphi''(b-a)}{\varphi(b-a)}+\mathcal{O}(\varepsilon^{3}).
\end{equation}

Now for $f^{(0)}$ we have
\begin{equation}
  \label{eq:61}
  f^{(0)}=\frac{1}{ab}\sum_{n=1}^{\infty} \frac{\tilde{\chi}_{n}\left(n\varepsilon
      \varphi(b-a),\varepsilon\right) }{n^{3}\varphi(b-a)^{2}},
\end{equation}
where $\tilde{\chi}_{n}(z,\varepsilon)$  is defined as
\begin{equation}
  \label{eq:62}
  \tilde{\chi}_{n}(z,\varepsilon)\equiv \frac{z^{2} e^{-z}
    H_{n}(\varepsilon)}{\left(1-e^{-z\left(1+\alpha_{-}(\varepsilon)\right)}\right)
    \left(1-e^{-z\left(1+\alpha_{+}(\varepsilon)\right)}\right)}.
\end{equation}
It is important to include $H_{n}(\varepsilon)$, since $H_{n}(\varepsilon)$ contains decaying
exponents in $n$.

Taking the derivatives of $f^{(0)}$ in $\varepsilon$ we get
\begin{multline}
  \label{eq:63}
  \left(\frac{d}{d\varepsilon}\right)^{2} f^{(0)}=\frac{1}{ab\varphi(b-a)} \sum_{n=1}^{\infty}
  \frac{1}{n^{3}}\left[
    \left.\tilde{\chi}''_{n,\varepsilon\varepsilon}\left(n\varepsilon\varphi(b-a),\varepsilon\right)\right|_{\varepsilon=0}+
    2n\varphi(b-a)\left.\tilde{\chi}''_{n,z\varepsilon}\left(n\varepsilon\varphi(b-a),\varepsilon\right)\right|_{\varepsilon=0}+\right.\\\left.
      n^{2}\left(\varphi(b-a)\right)^{2}\tilde{\chi}''_{n,zz}\left(n\varepsilon\varphi(b-a),\varepsilon\right)  \right]+
    \mathcal{O}(\varepsilon \ln \varepsilon)=\\
    =\frac{1}{ab\varphi(b-a)}\left[\sum_{n=1}^{\infty} \frac{1}{n^{3}}
      \tilde{\chi}''_{n,\varepsilon\varepsilon}(0,0)+2\varphi(b-a)\sum_{n=1}^{\infty}\frac{1}{n^{2}}\tilde{\chi}''_{n,z\varepsilon}(0,0)\right]+
    \\
    +\frac{1}{ab}\sum_{n=1}^{\infty}\frac{H_{n}}{n}\chi''\left(n\varepsilon\varphi(b-a)\right)+\mathcal{O}(\varepsilon\ln \varepsilon),
\end{multline}
where the function $\chi(z)$ is the same as in the equation \eqref{eq:39}
\begin{equation}
  \label{eq:64}
  \chi(z)=\frac{z^{2} e^{-z}}{\left(1-e^{-z}\right)^{2}}
\end{equation}
and
\begin{equation}
  \label{eq:65}
  H_{n}=\left(1-e^{-nb\varphi(b-a)}\right) \left(1-e^{-na\varphi(b-a)}\right).
\end{equation}

The last sum in the last line of the equation \eqref{eq:63} is the same as in the section
\ref{sec:gener-other-doma}, thus we can compute it:
\begin{multline}
  \label{eq:66}
  \frac{1}{ab}\sum_{n=1}^{\infty}\frac{H_{n}}{n}\chi''\left(n\varepsilon \varphi(b-a)\right)=
  \frac{1}{ab}\left(\frac{1}{6}\ln \varepsilon +\frac{1}{6}\ln \varphi(b-a)+ \int_{0}^{\infty}
    e^{-z}Q(z) \dz+\chi''(0)\sum_{n=1}^{\infty}\frac{H_{n}-1}{n}\right)=\\
  =\frac{1}{ab}\left(\frac{1}{6}\ln \varepsilon +\frac{1}{6}\ln \varphi(b-a)+ \int_{0}^{\infty}
    e^{-z}Q(z) \dz+\frac{1}{6}\ln\left(\frac{\left(1-e^{a\varphi(b-a)}\right)\left(1-e^{a\varphi(b-a)}\right)}{1-e^{(a+b)\varphi(b-a)}}\right)\right),
\end{multline}
where $Q(z)$ was defined in the equation \eqref{eq:85}.

Thus the contribution of $f^{(0)}$ to the asymptotic expansion \eqref{eq:26} consists of the term 
\begin{equation}
  \label{eq:87}
f^{(0)}_{3}=\frac{1}{ab\varphi(b-a)}\left[\sum_{n=1}^{\infty} \frac{1}{n^{3}}
  \tilde{\chi}''_{n,\varepsilon\varepsilon}(0,0)+2\varphi(b-a)\sum_{n=1}^{\infty}\frac{1}{n^{2}}\tilde{\chi}''_{n,z\varepsilon}(0,0)\right],
\end{equation}
that is cumbersome but does not depend on $\varepsilon$ and can be computed explicitly and the
expression \eqref{eq:66}, which gives us logarithmic behavior in $\varepsilon$.

Now we need to consider other terms in the expression \eqref{eq:55}. For $f^{+}$ we have
\begin{equation}
  \label{eq:67}
  f^{+}=\frac{1}{ab}\sum_{n=1}^{\infty}\frac{1}{n} e^{-n\varepsilon\varphi(b-a)} \varepsilon
  \frac{1-e^{-nbr_{-}'(0)}}{1-e^{-n\varepsilon r_{-}'(0)}}\varepsilon\sum_{j=0}^{M-1}\Psi^{+}_{n}(j\varepsilon).
\end{equation}
The sum over $j$ can be approximated by an integral using Euler-Maclaurin formula as
\begin{equation}
  \label{eq:68}
  \varepsilon\sum_{j=0}^{M-1}\Psi^{+}_{n}(j\varepsilon) = \int_{0}^{a}\Psi^{+}_{n}(x) \dx -
  \frac{\varepsilon}{2}\left(\Psi^{+}_{n}(a)-\Psi^{+}_{n}(0)\right)+
  \frac{\varepsilon^{2}}{12}\left(\Psi^{+'}_{n}(a)-\Psi^{+}_{n}(0)\right)+
  \mathcal{O}\left(\varepsilon^{m}\Psi^{+(m-1)}_{n}\right).
\end{equation}
Since $r(x)$ is an analytic function and $r(0)=0$ we can write $r(x)=xr'(0)+x^{2}\zeta(x)$, so 
\begin{equation}
  \label{eq:83}
\Psi_{n}(x)=e^{-nr'(0)x}\left(e^{-n\left(r(x)-xr'(0)\right)}-1\right)=e^{-nxr'(0)}\left(e^{-nx^{2}\zeta(x)}-1\right)  
\end{equation}
and we see that $\Psi_{n}(0)=0$ and $\Psi_{n}'(0)=0$.
Than in the limit $n\to\infty$ we have
\begin{equation}
  \label{eq:69}
  \int_{0}^{c}\Psi_{n}(x)
  \dx \xrightarrow[n\to\infty]{}\int_{0}^{c}e^{-nxr'(0)}\left(-nx^{2}\zeta(x)+\dots\right)\dx\sim \frac{1}{n^{2}}.
\end{equation}
The derivatives of $\Psi$ behave as $\Psi_{n}^{(m)}(0)\sim n^{m-1}$, since the derivatives should
fall on the bracket in the expression \eqref{eq:83} for the value to be non-zero, and then extra
power of $x$ is produced, which in turn should be differentiated. So the contribution from
$\varepsilon^{m}\Psi_{n}^{(m-1)}(0)$ to $f^{+}$ gives us
\begin{equation}
  \label{eq:70}
  \sum_{n=1}^{\infty} \frac{1}{n} e^{-n\varepsilon\varphi(b-a)}
  \frac{1-e^{-nbr_{-}'(0)}}{1-e^{-n\varepsilon r_{-}'(0)}} \varepsilon^{m+1}\Psi_{n}^{(m-1)}(0)\sim
  \sum_{n=1}^{\infty} \frac{1}{n^{4}} e^{-n\varepsilon\varphi(b-a)}
  \frac{1-e^{-nbr_{-}'(0)}}{1-e^{-n\varepsilon r_{-}'(0)}} (\varepsilon n )^{m+1}.
\end{equation}
For the second derivative in $\varepsilon$ we thus see that main contribution behaves as
$\sum_{n=1}\frac{1}{n^{2}}e^{-n\varepsilon\varphi(b-a)}
  (n\varepsilon)^{m+1}\xrightarrow[\varepsilon\to 0]{}0$
if $m\geq 3$. For the terms with $m=0,1,2$ we need to
take the limit $\varepsilon\to 0$. Doing so, we obtain
\begin{multline}
  \label{eq:71}
  f^{+}_{3}\equiv\left.\left(\frac{d}{d\varepsilon}\right)^{2} f^{+}\right|_{\varepsilon=0} =\\
  \left.\left(\frac{d}{d\varepsilon}\right)^{2}\frac{1}{ab}\sum_{n=1}^{\infty} \frac{1}{n}
  e^{-n\varepsilon\varphi(b-a)} \frac{1-e^{-nbr_{-}'(0)}}{1-e^{-n\varepsilon r_{-}'(0)}}
  \varepsilon\left(\int_{0}^{a} \Psi_{n}^{+}(x) \dx -\varepsilon \Psi_{n}^{+}(a) +\varepsilon^{2}\Psi^{+'}_{n}(0)\right)\right|_{\varepsilon=0}.
\end{multline}
Note that $\frac{\varepsilon}{1-\exp(-n\varepsilon r'_{-}(0))}$ is analytic in $\varepsilon$. 
The derivative can be computed explicitly, but the expression is too long to present it here.

The contribution of $f^{-}$ is computed in exactly the same way:
\begin{multline}
  \label{eq:72}
   f^{-}_{3}\equiv \left.\left(\frac{d}{d\varepsilon}\right)^{2} f^{-}\right|_{\varepsilon=0} =\\
  \left.\left(\frac{d}{d\varepsilon}\right)^{2}\frac{1}{ab}\sum_{n=1}^{\infty} \frac{1}{n}
  e^{-n\varepsilon\varphi(b-a)} \frac{1-e^{-nar_{+}'(0)}}{1-e^{-n\varepsilon r_{+}'(0)}}
 \varepsilon\left(\int_{0}^{b} \Psi_{n}^{-}(x) \dx -\varepsilon \Psi_{n}^{-}(a) +\varepsilon^{2}\Psi^{-'}_{n}(0)\right)\right|_{\varepsilon=0}.
\end{multline}
At last, we need to consider
$f^{\times}=\frac{1}{ab}\sum_{n=1}^{\infty}\frac{1}{n}e^{-n\varepsilon\varphi(b-a)}\varepsilon\sum_{i=0}^{N-1}\Psi_{n}^{-}(i\varepsilon)
\sum_{j=0}^{M-1}\Psi_{n}^{+}(j\varepsilon)$. Analyzing the contributions of the derivatives
$\Psi_{n}^{(m)}(0)$, similarly to the discussion before the formula \eqref{eq:70}, we see that
$\varepsilon^{m}\Psi_{n}^{(m-1)}(0)\sim \frac{1}{n^{2}}$. Thus
\[\sum_{n=1}^{\infty}\frac{1}{n}e^{-n\varepsilon\varphi(b-a)}\frac{1}{n^{2}}\varepsilon^{m}\Psi_{n}^{(m-1)}(0)\sim
\sum_{n=1}^{\infty}\frac{1}{n^{5}}e^{-n\varepsilon\varphi(b-a)}(\varepsilon n)^{m},\] and these terms
give zero contribution after the derivative $\left(\frac{d}{d\varepsilon}\right)^{2}$ for
$\varepsilon\to 0 $ and $m\geq 3$. The same holds for all the contributions with the derivatives of
$\Psi$ in the both sums. In the terms with $m<3$ we can just take a double derivative over
$\varepsilon$ and limit $\varepsilon\to 0$ and obtain
\begin{multline}
  \label{eq:73}
   f^{\times}_{3}\equiv \left.\left(\frac{d}{d\varepsilon}\right)^{2}
    f^{\times}\right|_{\varepsilon=0}=\left.\frac{1}{ab}\sum_{n=1}^{\infty}\frac{1}{n}
  \left(\frac{d}{d\varepsilon^{2}}\right)^{2}\left(\int_{0}^{b}\Psi^{-}_{n}(x)\dx
    -\frac{\varepsilon}{2}\Psi^{-}_{n}(b)+\frac{\varepsilon^{2}}{12}\Psi^{-'}_{n}(b)\right)\cdot\right.\\
\cdot\left.\left(\int_{0}^{a}\Psi^{+}_{n}(x)\dx-\frac{\varepsilon}{2}\Psi^{+}_{n}(a)+
    \frac{\varepsilon^{2}}{12}\Psi^{+}_{n}(a)\right)\right|_{\varepsilon=0}.
\end{multline}
Again, this expression is easily computed explicitly, but is too cumbersome to present it here.

At last we can write the asymptotic expansion coefficients in the present case:
\begin{equation}
  \label{eq:84}
  f_{0}= -\frac{1}{ab}\int_{0}^{b}\int_{0}^{a}\ln
  \left(1-e^{-\int_{-y}^{z}\varphi\left(b-a+x\right)\dx} \right)\dz\dy,
\end{equation}
\begin{equation}
  \label{eq:88}
  f_{1}=0,
\end{equation}
\begin{equation}
  \label{eq:89}
  f_{2}=\frac{1}{ab}\frac{1}{12}
\end{equation}
and
\begin{multline}
  \label{eq:90}
  f_{3}=f^{(0)}_{3}+f^{+}_{3}+f^{-}_{3}+f^{\times}_{3}+
  \frac{1}{ab}\frac{1}{12}\ln\left(\frac{\varphi(b-a)\left(1-e^{a\varphi(b-a)}\right)\left(1-e^{a\varphi(b-a)}\right)}{1-e^{(a+b)\varphi(b-a)}}\right)+\\
  +\frac{1}{ab}\left(\frac{1}{8} +\frac{1}{2} \int_{0}^{\infty}
    e^{-z}q(z) \dz\right),
\end{multline}
where in the first line we have the contributions that depend on the choice of the weight function
$\varphi$ and in the second line -- the universal contribution that appears in the case
$\varphi\equiv 1$ and in the box of finite height.

\subsection{Numerical checks}
\label{sec:numerical-checks-nonuniform}
Here we present results of comparison of numerical values of the expansion coefficients
$f_{0},f_{1},f_{2},f_{3}$ that were obtained from analytical formulas
\eqref{eq:84},\eqref{eq:88},\eqref{eq:89},\eqref{eq:90} using Mathematica computer algebra system
with the results of fitting exact numeric values of free energy for $\frac{1}{\varepsilon}=2\dots
200$
by the formula $f=f_{0}+
f_{1}\varepsilon+
f_{2}\varepsilon^{2}\ln\varepsilon+f_{3}\varepsilon^{2}+f_{4}\varepsilon^{3}+f_{5}\varepsilon^{4}$. 

In the table we present results for various functions $\varphi(z)$ and values of $a,b$.
\begin{table}[htpb]
\begin{center}
\begin{tabular}{c|cccccc}
  & $\varphi(z)$ & $a,b$ & $f_{0}$ &  $f_{1}$ &  $12ab f_{2}$ &  $f_{3}$ \\
 \hline
 analytic & $(2+\cos(z))/3$ & 1,3 & $0.472206688$ & $0$ & $1$&$-0.043883000$\\
 numeric & $(2+\cos(z))/3$ & 1,3 &  $0.472206693$ & $0.000000730$ & $1.000730689$&$-0.043827958$\\
  \hline
 analytic & $(2+\cos(z))/3$ & 2,3 & $0.235922467$ & $0$ & $1$&$-0.030398000$\\
 numeric & $(2+\cos(z))/3$ & 2,3 &  $0.235922467$ & $0.000000044$ & $1.000088528$&$-0.030394694$\\
\hline                                                
 analytic & $a+z/2$ & 1,3 & $0.097288596$ & $0$ & $1$&$-0.033688300$\\
 numeric & $a+z/2$ & 1,3 &  $0.097288593$ & $0.000000848$ & $1.000848267$&$-0.033624441$\\
\hline                                                
 analytic & $a+z/2$ & 2,3 & $0.032804447$ & $0$ & $1$&$-0.015094000$\\
 numeric & $a+z/2$ & 2,3 &  $0.032804447$ & $-0.000000002$ & $0.999996808$&$-0.015094162$\\
\end{tabular}
\caption{Values of the fitted parameters and analytically computed expansion coefficients for
  various weight functions $\varphi$ and values of $a,b$.
}
\label{tttt1}
\end{center}
\end{table}

We see very good coincidence between values, obtained numerically and analytically.

\section*{Conclusion and outlook}
\label{sec:conclusion}

In the present paper we computed the asymptotic expansion of the free
energy in the dimer model on a hexagonal domain of the hexagonal
lattice, on a hexagon with a side of infinite length and in the model
with coordinate-dependent Boltzmann weights. We presented numerical
results supporting our computations. We've discussed the physical
meaning of the expansion coefficients and argued that our results
support the identification of the scaling behavior of the dimer model
with the Gaussian free field theory.

In further work we will study the connection of the expansion
coefficients with the spectral properties of Dirac operator on the
non-frozen domain and study the universality of our results. We will
also present the computations of the logarithmic correction to the free
energy from the geometry of the ``limit shape''.

\section*{Acknowledgments}
\label{sec:acknowledgements}
We are grateful to professor Nikolai Reshetikhin for his attention to
this work. We thank Pavel Belov for useful discussions and general
support. We thank Mikhail Vyazovsky for pointing out a mistake in the
first version of the paper. We thank Istvan Prause for the reference
\cite{allegra2015exact} and Vadim Gorin for a discussion on geometry
of the limit shape.

We thank the organizers and participants of the conference MQFT-2018
for the opportunity to present our results and useful discussions.

This research is supported by RFBR grant No. 18-01-00916.

\bibliographystyle{utphys}
\bibliography{listing,bibliography,dimers}{} 
\end{document}